\documentclass[rnote]{aa}
\usepackage{graphicx}
\usepackage{txfonts}
\def\rxj{RX\,J1554.2+2721}
\def\kmps{km\,s$^{-1}$}
\begin{document}
\title{A 110\,MG cyclotron harmonic in the optical 
spectrum of \rxj}

\author{A.D. Schwope\inst{1}
\and
M.R. Schreiber\inst{1}
\and
P. Szkody\inst{2}}

\offprints{A. Schwope}

\institute{Astrophysikalisches Institut Potsdam, An der Sternwarte 16,
           14482 Potsdam, Germany
\and
Dept. of Astronomy, Box 351580, U of Washington, Seattle, WA 98195, USA}

\date{Received 22 December 2005; accepted 28 February 2006}

\abstract{
We report the detection of a 110\,MG cyclotron harmonic in the
  SDSS-spectrum of the magnetic cataclysmic variable (MCV) \rxj. This feature
  was noted earlier by others but remained unexplained. The wide spectral
  coverage of the new spectrum together with the earlier detection of a Zeeman
  split Ly$\alpha$ line in a field of 144 MG makes
  the identification almost unambiguous. 
  We propose to explain the non-conforming
  UV-optical photospheric temperature of the white dwarf
  by an as yet unobserved cyclotron component
  in the ultraviolet which also could significantly contribute to the overall
  energy balance of the accretion process. 
\keywords{Starts: cataclysmic variables -- Stars: magnetic fields --
  Stars:individual: \rxj -- Radiation mechanisms: thermal} }
\titlerunning{A 110\,MG cyclotron harmonic in \rxj}
\maketitle
%

\section{Introduction}
AM Herculis stars, also termed polars, form the sub-group of
cataclysmic variable stars hosting a strongly magnetic white dwarf primary as
accretor. The field prevents the formation of an accretion disk and keeps both
stars in synchronous rotation. Accretion happens via Roche-lobe overflow and
streams/curtains down to the polar regions of the white dwarf. The accretion
plasma, typical temperatures are $kT \simeq 10$\,keV, cools via thermal plasma
radiation in the X-ray domain and via cyclotron radiation. Depending on
the field strength in the accretion region, which is between 10 and 60 MG for
most of the polars, the cyclotron spectrum emerges at
IR or optical wavelength. Cyclotron radiation in the UV is equally possible
but observed only in a few cases due to the sparcity of high-field systems
(see Rosen et al.~2001, G\"ansicke et al.~2001, and Nogami et al.~2002 for
results on QS Tel, AR UMa, and UZ For, respectively).

\rxj\ was identified as a CV in the Hamburg Quasar Survey (Jiang et al.~2000)
and independently by Tovmassian et al.~(2001, henceforth TEA01) as the
optical counterpart of a soft 
RASS source. Its magnetic nature was suggested by Tovmassian et al.~who
also tentatively identified low frequency flux variations as cyclotron
harmonics in a field of $\sim$30\,MG. They also recognized its period in the
CV period gap. Thorstensen \& Fenton (2002, henceforth TF02) determined an
accurate orbital 
period of $P_{\rm orb} = 151.865$\,min right in the centre of the period gap
and 
estimated the distance to the system on the basis of spectral features of the
secondary star to be roughly 210\,pc. They also noted a pronounced
double-humped structure of the I-band light curve due to ellipsoidal
modulation of the secondary and a spectral hump of possible magnetic origin at
around 5000\,\AA. Finally, G\"{a}nsicke et al.~(2004, henceforth GEA04)
reported the detection 
of Zeeman split Ly$\alpha$ absorption lines in a HST-STIS snapshot
spectrum. The UV data were successfully modeled with a centered dipole model
with polar field strength 144\,MG and photospheric temperature of 17500\,K,
thus making \rxj\ only the third polar with a field strength in excess of
100\,MG. The combined optical/UV spectral energy distribution, however,
suggested a significantly higher temperature of about 23000\,K.

\section{A new SDSS spectrum of \rxj}
\begin{figure}[t]
\resizebox{\hsize}{!}{\includegraphics[clip=]{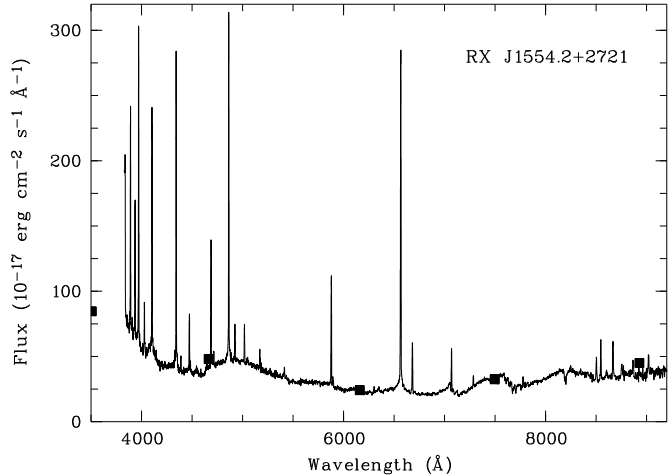}}
\caption{SDSS-spectrum of \rxj\ obtained May 8, 2005. Also shown is the
  $ugriz$ photometry obtained June 22, 2003.} 
\label{f:spec}
\end{figure}

Our review on \rxj\ is motivated by a single spectrum 
taken by the SDSS on May 8, 2005. The spectrum of the object with designation
SDSS J155412.33+272152.4, obtained  with an exposure time of 46 min,
is reproduced in Fig.~\ref{f:spec}. Also shown in the Figure is the SDSS
photometry obtained June 22, 2003, with $u=17\fm60, g=17\fm55, r= 17\fm70, i=
16\fm95, \mbox{ and } z = 16\fm19$, which shoes very little deviation from the
spectroscopy. For details of the SDSS project the reader is refered to 
Adelman-McCarthy et al.~(2006),
Fukugita et al.~(1996),
Gunn et al.~(2006), and
York et al.~(2000).

The accumulated phase uncertainty at the time of the spectral observations,
i.e.~after 13456 cycles according to the
ephemeris of TF02 is $\Delta\phi = 0.75$, hence we cannot assign a proper 
orbital phase to the spectrum. 
The main features of the spectrum are unchanged with
respect to the observations by TF02 and TEA01, the prevalence of the M4
secondary in the red spectral range, a hump centered on $\sim$5000\,\AA,
prominent high-state emission lines of H and He and a spectral upturn at the
blue end.  

The new SDSS spectrum has a continuum flux level very similar to the March
2001 low-state spectrum by TF02 (their Fig.~3b) but a very different emission
line spectrum containing even lines of ionized Helium. The low-state spectrum
of TF02 has $\sim$20\% 
less flux in the range 5000--7000\,\AA, 10\% more flux at 4600\,\AA\ and
matches exactly at 7600\,\AA, where mainly the secondary contributes. 
The pronounced difference in the line spectrum between TF02 low state
and SDSS suggests the
presence of a soft X-ray ionizing source at the time, when the SDSS
spectrum was taken, hence a certain level of
accretion. The SDSS-spectrum was thus not obtained in a low accretion state
although the continuum flux level might be regarded as indicative of a low
state. 
 
The May 2001 high-state spectrum of TF02 has a very similar line spectrum
compared to the SDSS spectrum but an enhanced continuum flux by
about 20--30\%. The red part of the high-state spectrum by TEA01 (their
Fig.~4b) 
matches exactly the new SDSS spectrum while the blue part of their spectrum
taken one night later is again about 25\% brighter.
Taken together, neither 
the orbital variability nor the rather frequent changes between high and low
states seem to have a pronounced influence on the system's brightness
(although one must admit that high-state photometry in the blue part
of the spectrum is missing).  

The atomic emission lines in the SDSS-spectrum seem to be unusually sharp and
peaked. They can be fit with the superposition of a broad and a narrow Gaussian
line with FWHM $\sim$2000\,\kmps\ and 
250--300\,\kmps (after deconvolution with the instrumental profile). 
These narrow lines appear wider than the narrow emission
lines in other polars which are of reprocessed origin from the secondary (FWHM
$\sim$100\,\kmps) but velocity smearing is likely be responsible for the
relative large width. 
The broad line clearly indicates an origin in an accretion stream.

The very feature which makes the spectrum of \rxj\ outstanding is the hump
centered 
on $\sim$ 4950\,\AA, which we regard as a cyclotron emission
line, either the cyclotron fundamental or a low cyclotron harmonic. 
This feature was
present in the spectra of TEA01 and TF02 too but didn't show up so prominently
due to the shorter wavelength coverage and blending with atomic emission
lines. 

\subsection{The cyclotron harmonic at 4950\,\AA}
In oder to isolate the cyclotron component we subtracted a scaled template
M-dwarf spectrum and the 20000\,K magnetic white dwarf model spectrum by
GEA04 adjusted to the flux level around 4600\,\AA. For subtraction of the
template M-dwarf we initially used the same spectrum of Gl447 (M4+) as
TF02. The SDSS spectrum required a scaling parameter 20\% different from TF02
which led to a slightly larger distance estimate, 235\,pc instead of 210\,pc. 
We finally used a spectrum of GJ\,268 (M4.5) available to us comprising a
wavelength range similar to that of the SDSS-spectrum.
The difference spectrum is regarded to mainly consist of cyclotron radiation 
but might be affected by Zeeman absorption and might have additional
contributions from the recombination continuum. On the basis of just one new
spectrum the latter two components cannot be quantified. 

\begin{figure}[t]
\resizebox{\hsize}{!}{\includegraphics[clip=]{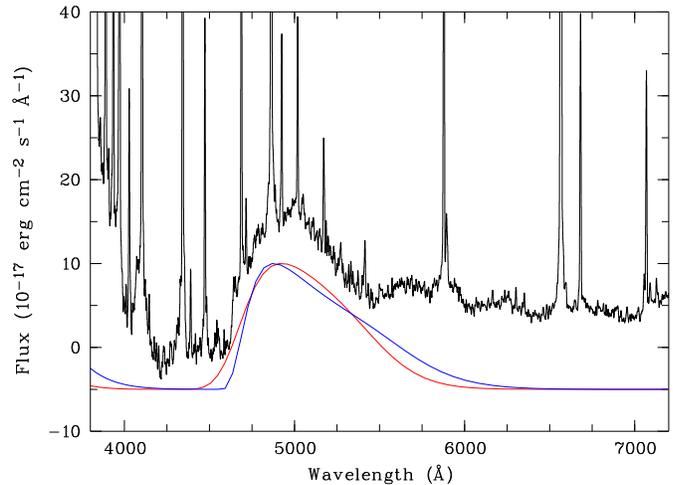}}
\caption{Cyclotron spectrum of \rxj\ computed by sutraction of suitably scaled
  M-dwarf and white dwarf spectra. The model spectra for 5 and 10\,keV are
  plotted with an offset of $-$5 units.}
\label{f:ofit}
\end{figure}

The difference spectrum is reproduced in Fig.~\ref{f:ofit}.
It shows a single cyclotron harmonic rising sharply at 4600\,\AA\  and
dropping of 
at $\sim$5450\,\AA. The red continuum has more flux than the blue continuum
between 4150--4600\,\AA. The use of a cooler 15000\,K white dwarf instead of
the 20000\,K does not cure the problem. 
Either there is an emission component in
the red part of the spectrum or the blue part is affected by Zeeman
absorption. Again, on the basis of just one spectrum we cannot decide between
the alternatives but our main conclusions are not affected by these
uncertainties.

The lack of any neighboring cyclotron harmonic excludes any
low-field/high-harmonic interpretation of the observed structure. The 
144\,MG field determined by GEA04 suggests an identification with either the
cyclotron fundamental or the second harmonic. If it would be the fundamental
then the derived field strength would be 220\,MG. This exceeds the implied
polar field 
strength by a large amount and is regarded as the less likely, although not
impossible interpretation. If for instance the photospheric spectrum would be
obtained under an unfavourable viewing geometry or the field structure
deviate from the centered dipole model, as many polars do, the assumed polar
field strength can be very much different from the real value. However, unless
other evidence is available the identification with the second harmonic seems
plausible. 

We have modeled the harmonic with isothermal plasma cyclotron radiation using
the code described in previous papers (Schwope et al.~1990). Two model curves
with slightly different field strength $B$, plasma temperature $kT$, aspect
angle $\theta$, but the same density parameter $\Lambda$ are also
shown in Fig.~\ref{f:ofit}. The density parameter $\Lambda = 
l \omega_p^2 / c\omega_c = 5\times 10^4
(l/10^5\mbox{ cm}) (n_e/10^16\mbox{ cm}^-3) (110\mbox{ MG})/B)$
essentially determines where the spectrum turns from the low-frequency
optically thick to the high-frequency optically thin part.
The angle $\theta$ is the angle between the line
of sight and the magnetic field. 
The narrower line was calculated for $kT=5$\,keV,
$B=108$\,MG, $\theta = 60\degr$, $\log\Lambda = 1$, the broader line for
10\,keV, 113\,MG, and $70\degr$, and $\log\Lambda = 1$. 
The width of individual cyclotron lines is a function of
$\theta$ and $T$, but a scatter in these values and, in particular, 
a scatter in $B$ may further broaden individual lines. 
We cannot discern between these different 
broadening mechanisms and the models were calculated with the most simple
assumptions, a homogeneous plasma with $\theta$ adjusted to match the observed
width of the line.


\section{Discussion}
\subsection{The UV-optical spectral energy distribution}
There is evidence that the one single cyclotron harmonic is not fully 
representative of the cyclotron spectrum of \rxj\ in the main accretion
region. The value of the density parameter
used for the fits shown in Fig.~\ref{f:ofit} was
extremely low, $\log\Lambda = 1$. This value is comparable to or even lower
than those in 
the recently discovered LARPs (Low Accretion Rate Polars, Schwope et al.~2002,
Schmidt et al.~2005) with permanent low accretion rates probably fed from a
wind. \rxj\ clearly does not belong to this elusive class of objects. It was a
RASS source, it shows prominent atomic emission lines of highly ionized
species, and high-state Doppler tomograms indicate Roche-lobe overflow
(TEA01). 

\begin{figure}[t]
\resizebox{\hsize}{!}{\includegraphics[clip=]{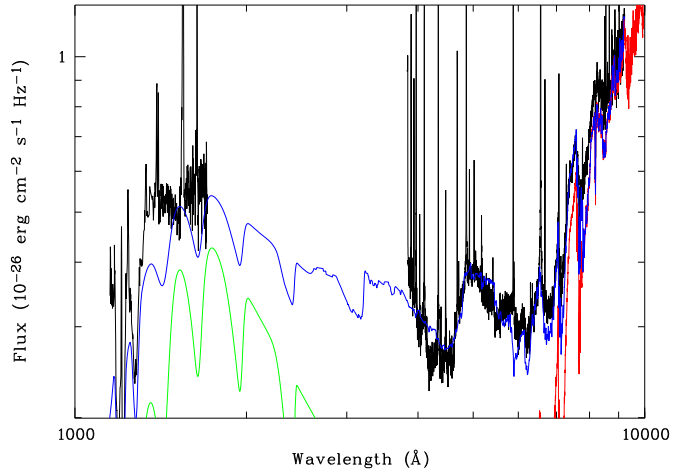}}
\caption{Combined optical/ultraviolet spectrum of \rxj\ using the STIS
  spectrum of GEA04 and the new SDSS spectrum. The composite model spectrum
  shown in blue consists of 
  a suitably scaled M-star spectrum (red), two cyclotron model spectra (green)
  and the 
  20000\,K white dwarf model of GEA04.} 
\label{f:sed}
\end{figure}

While \rxj\ seems to belong to the normal accreting polars displaying high and
low accretion states, the absence of a pronounced increase in the continuum
when switching from high to low states is puzzling. One distinct possibility is
that the main accretion region is continuously hidden from our view due to an
unfavourable geometry. In this model the observed cylotron harmonic would
belong to a secondary region with much lower accretion rate (as in VV Pup or
UZ For, see Wickramasinghe et al.~1989, Schwope \& Beuermann 1997, Schwope et
al.~1990). This scenario can be tested by a decent X-ray
observation in a high accretion state (indicated by the presence of strong
atomic emission lines) which should result in a very low X-ray signal.

Here we discuss shortly an alternative model.
Beuermann (2004) discussed a synthesized spectrum of AM Her composed of a
multi-temperature, multi-density plasma due to contributions from regions with
a wide range of specific mass accretion rates. Following his arguments we may
expect the major part of the cyclotron radiation emitted at higher harmonic
number in the ultraviolet. Assuming a similar scenario for \rxj\
we can give a rough estimate of those suspected contributions from the
ultraviolet-to-optical spectral energy distribution shown in
Fig.~\ref{f:sed}. The two observed spectra shown there are the STIS-spectrum
of GEA04 and the SDSS spectrum. GEA04 noticed that a higher temperature is
needed to match the SED than derived from just fitting the UV-spectrum. We
argue that the inclusion of a cyclotron component with spectral parameters that
one usually encounters in high-state polars solves the discrepancy. We
illustrate this with a composite spectrum consisting of the scaled M-dwarf
GJ\,268, the 20000\,K magnetic white dwarf model of GEA04 and two cyclotron
models, one with the extreme small density parameter $\log\Lambda = 1$, the
other with the more typical parameter $\log\Lambda = 5$. The synthetic
spectrum is not meant to be a proper fit of the data, since there are too  
many uncertainties in the spectral decomposition\footnote{
the main uncertainties are: -- the optical and UV spectra were not taken
simultaneously; -- there is no phase information for either of the two
spectra; 
-- we just used two cyclotron
models instead of a continuum} 
but merely an illustration of the
possible spectral composition. A proper composed synthetic spectrum
will probably explain the upturn at the blue end of the SDSS-spectrum, where
the third harmonic starts to rise. The integrated flux of our composite
cyclotron spectrum is $F=3\times 10^{-12}$\,erg\,cm$^{-2}$\,s$^{-1}$, a factor
30 higher than just the flux in the 2 harmonics and of the 
same order as the ROSAT
X-ray flux in the 0.1--2.4\,keV band (TEA01). Neglecting the UV-cyclotron
component would result in a heavily biased energy balance of the accretion
process for this particular source. 

\section{Conclusions}
We have presented an analysis of a new spectrum of \rxj\ obtained May 2005
during the SDSS. It shows a pronounced cyclotron hump which could be used to
measure the field strength in the accretion plasma and thus gives an
explanation to the spectral humps already noticed by TEA01 and TF02. 

At a field strength of 110\,MG one can access the very low harmonic numbers
usually hidden in the infrared or far infrared. This gives access to those
parts of a structured accretion region with very low specific mass accretion
rates. We propose to explain the non-conforming temperature
estimates from UV and optical spectroscopy by a missed cyclotron component in
the ultraviolet, which carries most of the cyclotron luminosity. 
Neglecting such a component could result in a heavily biased energy
balance of the accretion process towards thermal plasma radiation in
the X-ray regime. 
A full decomposition of the spectrum requires low-resolution
spectroscopy/spectrophotometry with full phase coverage, 
both in the optical and the ultraviolet. 

\begin{acknowledgements}
This project is supported by the Deutsches Zentrum f\"ur Luft- und Raumfahrt
(DLR) under contract no.~FKZ 50 OR 0404 (MRS) 
and by NSF grant AST 97-30792 (PS).

Funding for the Sloan Digital Sky Survey (SDSS) has been provided by the
Alfred P. Sloan Foundation, the Participating Institutions, the National
Aeronautics and Space Administration, the National Science Foundation, the
U.S. Department of Energy, the Japanese Monbukagakusho, and the Max Planck
Society. The SDSS Web site is http://www.sdss.org/. 

The SDSS is managed by the Astrophysical Research Consortium (ARC) for the
Participating Institutions. The Participating Institutions are The University
of Chicago, Fermilab, the Institute for Advanced Study, the Japan
Participation Group, The Johns Hopkins University, the Korean Scientist Group,
Los Alamos National Laboratory, the Max-Planck-Institute for Astronomy (MPIA),
the Max-Planck-Institute for Astrophysics (MPA), New Mexico State University,
University of Pittsburgh, University of Portsmouth, Princeton University, the
United States Naval Observatory, and the University of Washington. 

\end{acknowledgements}


\begin{thebibliography}{}
\bibitem[]{}
Adelman-McCarthy, J., et al., 2006, ApJS, in press (astro-ph/0507711)
\bibitem[]{}
Beuermann K., 2004, ASPC 315, 187
\bibitem[]{}
Fischer A., Beuermann K., 2001, A\&A 373, 211
\bibitem[]{}
Fukugita, M., Ichikawa, T., Gunn, J. E., Doi, M., Shimasaku, K., \&
    Schneider, D. P., 1996, AJ, 111, 1748
\bibitem[]{}
G\"{a}nsicke B., Schmidt, G., Jordan S., Szkody P., 2001, ApJ 555, 380
\bibitem[]{}
G\"{a}nsicke B., Jordan S., Beuermann K., de Martino D., Szkody P., Marsh T.,
Thorstensen J., 2004, ApJ 613, L141 (GEA04)
\bibitem[]{}
Gunn, J. E., et al. 2006, AJ, submitted
\bibitem[]{}
Nogami D., G\"ansicke B., Beuermann K., 2002, ASPC 261, 159
\bibitem[]{}
Jiang X.J., Engels D., Wei J.Y., Tesch F., Hu J.Y., 2000, A\&A 362, 263
\bibitem[]{}
Rosen S.R., Rainger J.F., Burleigh M.R., et al., 2001, MNRAS 322, 631
\bibitem[]{}
Schwope A.D., Beuermann K., 1997, AN 318, 111
\bibitem[]{}
Schwope A.D.,  Beuermann K., Thomas, H.-C., 1990, A\&A 230, 120
\bibitem[]{}
Schwope A.D.,  Brunner, H., Hambaryan V., Schwarz R., Staude A., Szokoly G.,
2002, ASPC 261, 102
\bibitem[]{}
Schmidt G.D., Szkody, P., Vanlandingham K.M., et al., 2005, ApJ 630, 1037
\bibitem[]{}
Thorstensen J., Fenton W.H., 2002, PASP 114, 74 (TF02)
\bibitem[]{}
Tovmassian G.H., Greiner J., Zharikov S.V., Echevarria J., Kniazev A., 2001,
A\&A 380, 504 (TEA01)
\bibitem[]{}
Wickramasinghe D.T., Ferrario L., Bailey, J., 1989, ApJ 342, L37
\bibitem[]{}
York, D. G., et al. 2000, AJ, 120, 1579
\end{thebibliography}
\end{document}